\documentstyle[aps,prl,preprint,epsfig,floats]{revtex}

\begin{document}

\bibliographystyle{prsty}

\draft 

\title{Coulomb Blockade Fluctuations \\ in Strongly Coupled Quantum Dots}

\author{S. M. Maurer, S. R. Patel and C. M. Marcus}

\address{Department of Physics, Stanford University, Stanford, California 94305}

\author{C. I. Duru\"oz and J. S. Harris, Jr.}

\address{Department of Electrical Engineering, Stanford University, Stanford,
California 94305}

\date{\today}

\maketitle

\begin{abstract}

Quantum fluctuations of Coulomb blockade are investigated as a function of the
coupling to reservoirs in semiconductor quantum dots. We use fluctuations in the
distance between peaks $\Delta N$ apart to characterize both the amplitude and
correlation of peak motion. For strong coupling, peak motion is greatly enhanced
at low temperature, but does not show the expected increase in peak-to-peak correlation.
These effects can lead to anomalous temperature dependence in the Coulomb
valleys, similar to behavior ascribed to Kondo physics.

\end{abstract}


The Coulomb blockade (CB) of tunneling through a confined island of charge (a
quantum dot) at low temperatures provides perhaps the clearest demonstration of
the interplay between electron-electron interactions and quantum effects in
reduced dimension. For weak tunneling from the dot to electronic reservoirs, CB
can be understood as a classical charging effect \cite{CB} modified by
mesoscopic fluctuations of the coupling of the dot to the leads
\cite{JalabertChangFolk}. As the coupling is increased, transport
becomes quite complicated and several effects appear that mix the influence of
interaction and quantum interference. These include elastic cotunneling, which
shows mesoscopic fluctuations on scales set by the charging energy
\cite{AleinerCot,SaraValleys}, and coherent enhancement of CB that is sensitive
to time-reversal symmetry breaking \cite{Aleiner,SaraFrying}. At very low
temperatures and strong tunneling, Kondo resonances between the spin of the dot
and the reservoirs further modify transport \cite{KondoTheory}, as observed in
recent experiments \cite{Goldhaber,SaraKondo}.

In this Letter, we report measurements of mesoscopic fluctuations of CB  peaks
as a function of tunneling strength in symmetrically coupled GaAs quantum dots.
Specifically, we examine fluctuations in height and position of CB peaks as a
function of coupling to reservoirs, temperature and dot size. The strong tunneling regime has
been investigated previously \cite{SaraValleys,Goldhaber,SaraKondo,Heinzel,Berman}, though
fluctuations of peak position have not been addressed.  Peak position statistics are of
particular interest because one expects quantum fluctuations to be greatly enhanced in the strong
tunneling regime as coherence and charging effects mix; one also expects a corresponding increase
in the number of peaks over which fluctuations are correlated
\cite{Aleiner,Kaminski}. Both of these effects arise from the fact that for strong tunneling a
large number $E_c /\Delta$ of quantum levels make comparable contributions to
conductance, where $E_c$ is the charging energy and $\Delta$ the mean quantum
level spacing. This contrasts the weak-tunneling regime, where low-temperature
transport is mediated by tunneling through the ground state only.

Our picture of strong tunneling in semiconductor quantum dots is based on recent
theory \cite{Aleiner,Kaminski,Brouwer} that extends inelastic cotunneling
\cite{FlensbergMatveev} and methods applicable to metallic grains \cite{Golubev}
by including elastic contributions. Refs.\ \cite{Aleiner,Kaminski} specifically
consider dots with asymmetric lead transmissions $T_1 \ll T_2 \le 1$ and so may
not be fully applicable here. Also, since neither mesoscopic fluctuations of
coupling strengths, nor Kondo-type resonances \cite{KondoTheory}, nor changes in
the energy spectrum upon addition of electrons \cite{Patel} are included, we do
not anticipate complete agreement between experiment and theory.

We find experimentally that CB peak motion is enhanced for strong tunneling (see
Fig.\ 1), as predicted, but that the number of correlated peaks is not enhanced
by the expected factor of $\sim E_c / \Delta$. This situation leads to
unanticipated experimental consequences, including anomalous (reversed)
temperature dependence of CB valleys, an effect that has been identified as a
signature of the Kondo effect in smaller devices \cite{Goldhaber,SaraKondo}.

We introduce a measure of peak fluctuations for an ensemble of CB peaks
as the standard deviation of the distance between peaks $\Delta N$ apart,
\[\sigma_p(\Delta N) = \langle{\langle (\tilde{V}_N - \tilde{V}_{N + \Delta
N})^2 \rangle_B}^{1/2}\rangle_N\] where $\tilde{V}_N = V_N - \langle V_N
\rangle_B$ is the position in gate voltage $V_N$ of the maximum conductance of
the $N^{th}$ peak minus its average over magnetic field, $\langle V_N
\rangle_B$.  The quantity $\sigma_p(\Delta N)$ generalizes the well-studied peak
spacing fluctuations, $\sigma_p(1)$
\cite{SamPeakSpacings,SivanSimmel,Berkovits}, and, unlike fluctuations of the
peak position itself, is not very sensitive to experimental drift \cite{drift}.
Noting that $\sigma_p(0) =0$, correlated peak motion  appears as a
reduction in $\sigma_p(\Delta N)$ for small $\Delta N$ whose width gives a measure of
the number of correlated peaks. For larger $\Delta N$, in chaotic or disordered dots
one would naively expect $\sigma_p(\Delta N) \propto \log{(\Delta N)}$ as long as $\Delta
N <  E_T / \Delta$, where $E_T \sim h v_F / \sqrt A$ is the Thouless energy
\cite{RMTAAK} for a ballistic dot of area $A$. Experimentally, we find
$\sigma_p(\Delta N)\propto \sqrt{\Delta N}$ using large data sets of 50 -- 100
consecutive peaks in the weak-tunneling regime \cite{PatelMaurer}; that is, we
do not observe long-range random-matrix-like correlations, presumably due to
changes in the addition spectrum as electrons are added \cite{Patel}. A related quantity,
$\sigma_g(\Delta N)$, can be similarly defined as the standard deviation of differences in peak
heights (i.e.\ conductance maxima) for peaks $\Delta N$ apart. Peak heights are not expected to
show long-range correlations. Rather, one expects  $\sigma_g(\Delta N)$ to saturate at
$\sqrt 2$ times the typical height fluctuations of a single peak.

Two scenarios for how strong tunneling might affect $\sigma_p(\Delta N)$ are
sketched in Figs.\ 1(d, e). In Fig.\ 1(d), fluctuations are larger for strong
tunneling, but the number of correlated peaks does not depend on 
tunneling strength. This reflects a picture in which a renormalized ratio of
$\Delta/E_c^*$ leads to larger peak motion, where $E_c^*$ is an effective
charging energy at strong tunneling \cite{renorm}. Figure 1(e) illustrates an alternative
cotunneling picture \cite{Aleiner,Kaminski} in which both fluctuations and
correlations increase with stronger tunneling. Although we expected Fig.\ 1(e)
to better describe our experiment, we find the data look more like Fig.\ 1(d).

Measurements are reported for two dots with areas $0.3\;\mu{\rm m}^2$ (small)
and $1.0\;\mu{\rm m}^2$ (large) [micrographs in Fig.\ 2], fabricated using CrAu
gates $90\;nm$ above a two-dimensional electron gas (2DEG) in a GaAs/AlGaAs
heterostructure. The 2DEG mean free path exceeds the device size, so
transport is ballistic within the dot. Charging energies $E_c = 300\;\mu eV$
(large dot) and $580\;\mu eV$ (small dot) were measured at weak tunneling from
the spacing and thermal width of CB peaks; quantum level spacings, $\Delta =
2 \pi \hbar^2 / m^* A \sim 7\;\mu e V$ (large dot) and $\sim 24\;\mu e V$ (small
dot), were estimated from the device area, excluding a depletion width of $\sim
70\;nm$. Measurements were made in a dilution refrigerator
using an ac voltage bias of $5\;\mu V$ at $13\;Hz$. The temperature $T$ refers
to the electron temperature, measured from weak-tunneling peak widths.

Figure 1 shows typical CB peaks as a function of gate voltage $V_g$
and magnetic field $B$, illustrating how peak position fluctuations increase
with tunneling strength. For weak tunneling, Fig.\ 1(a), the ratio of
fluctuations to average peak spacing is roughly $\sim \Delta/E_c$, consistent
with theory \cite{Fluctuations} and some \cite{SamPeakSpacings} but not all
\cite{SivanSimmel} previous experiments. Note, however, that
peak spacings gathered over $N$ \cite{SamPeakSpacings,SivanSimmel} need not have
the same statistics as either peak motions or peak spacings gathered in $B$. As
tunneling becomes stronger CB peaks show greatly enhanced motion as a function
of $B$, as seen in Fig.\ 1(b,c). Ensemble statistics of peak position and height fluctuations,
gathered from data sets similar to those in Fig.\ 1, are shown in Fig.\ 2. Each data set consists
of 15 -- 20 peaks, with fluctuations gathered over a range of magnetic fields
that include many flux quanta through the device ($\phi_0 / A \sim 4\;mT$ ($\sim
12\;mT$) for the large (small) dot) but remain below the field where the quantum
Hall effect appears.

Three features of these statistics are worth emphasizing.  First, peak position
fluctuations are seen to increase with stronger tunneling ( Fig.\ 2(a,c)). This is
important given the recent debate concerning the magnitude and origin of
fluctuations in peak spacing, $\sigma_p(1)$
\cite{SamPeakSpacings,SivanSimmel,Berkovits,Fluctuations}, and demonstrate that when
comparing fluctuations in different devices, tunneling strength must be taken into account.
Second, for the small dot peak position fluctuations actually decrease for the strongest
tunneling. This may be related to a similar theoretical result indicating  that capacitance
fluctuations in the valleys between CB peaks are nonmonotonic in coupling
strength \cite{Kaminski}.  Third, the number of correlated peaks does not depend on tunneling
strength. In other words, the curves in Fig.\ 2 more closely resemble Fig.\ 1(d) than Fig.\ 1(e).
This is further emphasized in Figs.\ 2 (b,d,f) which show the same curves,
normalized vertically by their values at large $\Delta N$. These
scaled curves, denoted $\tilde{\sigma}_p(\Delta N)$ and $\tilde{\sigma}_g(\Delta
N)$, each collapse onto a single curve with the same correlation length. This
scaling is not expected theoretically \cite{Aleiner,Kaminski}, but
experimentally appears valid for heights and positions at all measured
temperatures, from $45\;mK$ ($85\;mK$) to $300\;mK$ ($400\;mK$) in the small
(large) dot.

The absence of correlated peak motion for strong tunneling presumably reflects the fact that the
level spectrum of the dot changes as  electrons are added \cite{Patel}. It is known
from previous experiments
\cite{Patel,Duncan} that the spectrum near the Fermi surface is rearranged when the number of
electrons is changed by roughly 5 -- 8 for dots of this type and size. Since this number is
smaller than $E_c /
\Delta$, which is 40 (25) for the large (small) dot, these spectral changes rather than
elastic cotunneling set the number of correlated peaks.

The temperature dependences of the normalized CB fluctuations,
$\tilde{\sigma}_p(\Delta N)$ and $\tilde{\sigma}_g(\Delta N)$, averaged over
different tunneling strengths (a procedure justified by the scaling in Fig.\ 2)
are shown in Fig.\ 3.  The large dot shows the expected increased correlation at higher
temperatures ($k_BT > \Delta$). The small dot does not show increased correlations up to several
hundred millikelvin, probably due both to the effectively lower temperature (in units of
$\Delta$), and to the fact that fewer added electrons are needed to
rearrange the spectrum \cite{Patel}. The unnormalized amplitudes
of peak position fluctuations decrease with increasing temperature for both dots,
consistent with the expected dependence, $(k_BT/\Delta)^{-1/2}$ \cite{SamPeakSpacings}.

The fact that peak motion at strong tunneling is enhanced, but
correlations are not, can lead to situations in which neighboring CB peaks move
quite close together at certain magnetic fields. In this situation, as
the temperature is raised, peak fluctuations decrease and the two nearby peaks
move apart causing conductance in the valley between them to {\em decrease}. Several examples of
this are shown in Fig.\ 4. It is tempting to compare this behavior to recently reported
signatures of the Kondo effect in small quantum dots, also measured in the regime of strong
tunneling
\cite{Goldhaber,SaraKondo}. The anomalous temperature dependence of valley conductance that we
observe has a roughly logarithmic temperature dependence, as shown in the inset of Fig.\ 4(b),
similar to the Kondo data. However, unlike the even-odd character of the Kondo
effect, which leads to a strict alternation of anomalous and normal valleys, the
present ``peak wandering" effect can appear in adjacent valleys and can switch from one
valley to the next by small changes in field (of order $\phi_0/A$), as seen in
Fig.\ 4. For these anomalous valleys, finite bias measurements also show
similarities between the present strong-tunneling data and the Kondo data of
Refs. \cite{Goldhaber,SaraKondo}, including a zero-bias peak and a splitting of
the zero bias peak in a perpendicular $B$ field of order $1\;T$, consistent with
the value $50\;\mu V/T$ \cite{SaraKondo}, after averaging peaks and valleys over
gate voltage.

While the effects shown in Fig.\ 4 are not expected within the existing theory of cotunneling
\cite{Aleiner,Kaminski} it is not clear that they are related to the Kondo effect, despite the
similarities discussed above. On the other hand, a spin-dependent
strong-tunneling effect similar to a multilevel Kondo resonance \cite{Inoshita} could be
responsible for the enhanced but uncorrelated peak motion that leads to the
present anomalous temperature dependence of valleys, without requiring
even-odd alternation. This unexpected behavior demonstrates that the general
problem of transport through quantum dots, including interactions, strong
tunneling and spin effects continues to provide surprises
and challenges.

We thank I. L. Aleiner, S. M. Cronenwett, Y. Gefen, L. I. Glazman, L. P.
Kouwenhoven and D. R. Stewart for valuable discussions.  We acknowledge support
from the ARO under DAAH04-95-1-0331 and the NSF-PECASE under DMR 9629180-1 (Marcus
Group), and JSEP under DAAH04-94-G-0058 (Harris Group). SMM was supported by the
Hertz Foundation.

\begin{figure}[bth] \epsfxsize=\columnwidth \epsfbox{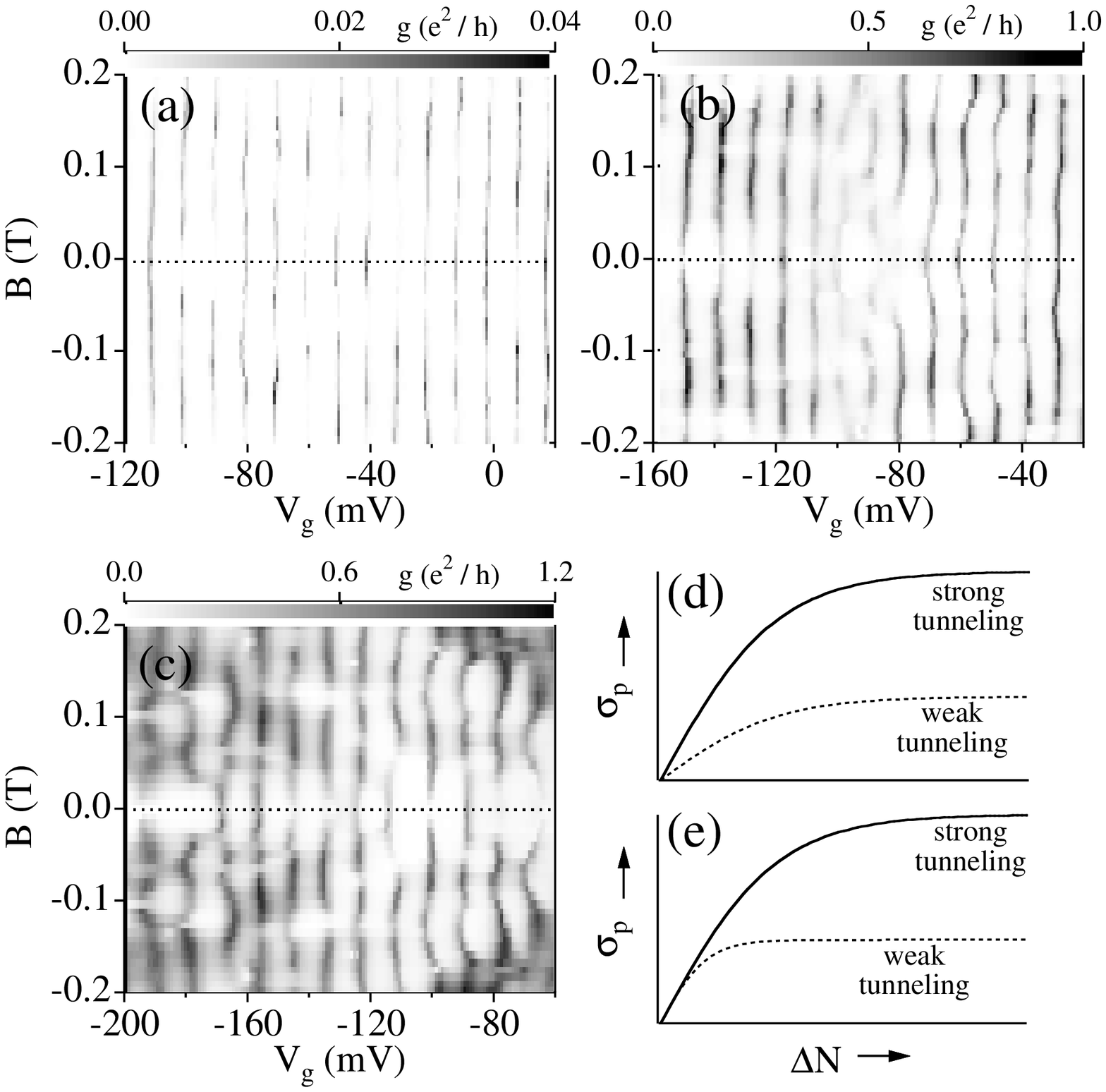}
\vspace{0.5 in}
\caption{(a, b, c) Grayscale plots of conductance in the
small dot ($0.3\;\mu{\rm m}^2$), showing typical CB peak fluctuations for weak
and strong tunneling.  $\langle g_{max} \rangle$ = 0.015 (a), 0.42 (b) and 0.61
(c). (d,e) Two scenarios for how fluctuations in peak separation
$\sigma_p(\Delta N)$ grow with the number of peaks $\Delta N$ in the weak and
strong tunneling regimes as described in the text.} \label{fig1} \end{figure}

\begin{figure}[bth] \epsfysize=5 in \epsfbox{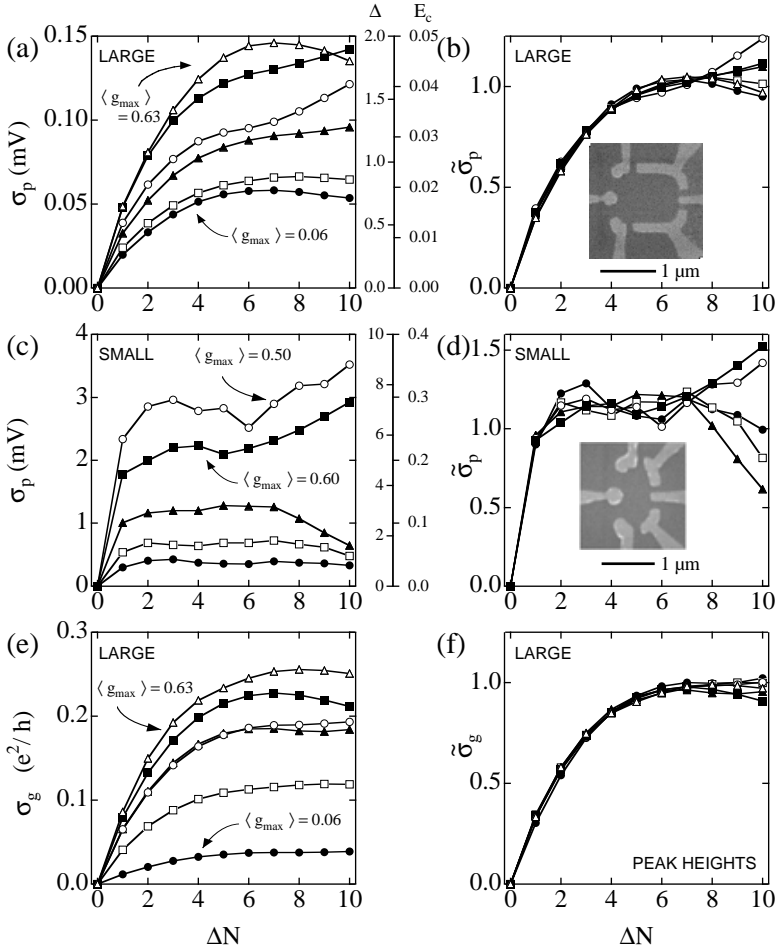}
\vspace{0.4 in}
\caption{(a) CB peak fluctuations $\sigma_p(\Delta N)$ for the large dot. From
bottom (filled circles) to top (open triangles), average peak heights are
0.062, 0.28, 0.49, 0.56, 0.60 and 0.63. Scales on right show
$\sigma_p(\Delta N)$ in units of mean level spacing $\Delta$ and charging energy
$E_c$.  (b) Same data as in (a) normalized by its value at large $\Delta
N$. Inset: micrograph of large ($1.0\;\mu{\rm m}^2$) dot. (c) CB peak
fluctuations  $\sigma_p(\Delta N)$ for the small dot. Average  peak
heights are 0.015 (closed circles), 0.32 (open squares), 0.42 (closed
triangles), 0.50 (open circles) and 0.60 (closed squares) $e^2/h$. Note that
fluctuations for $\langle g_{max} \rangle = 0.50$ are larger than for $0.60$.
(d) Normalized data from (c). Inset: micrograph of the small ($0.3\;\mu{\rm
m}^2$) dot. (e) Peak height fluctuations $\sigma_g(\Delta N)$ from the data in
(a). (f) Normalized peak height fluctuations, $\tilde{\sigma}_g(\Delta N)$, from
(e), has the same correlation as data in (b).} \label{fig2} \end{figure}

\begin{figure}[bth] \epsfxsize=\columnwidth \epsfbox{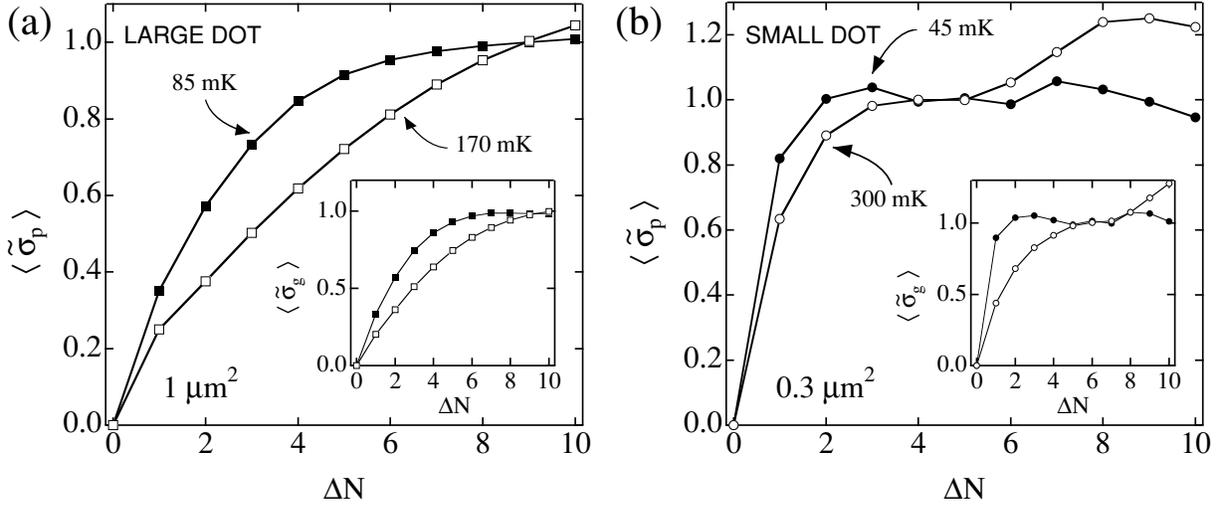}
\vspace{1 in}
\caption{(a) Temperature dependence of normalized CB peak fluctuations
$\tilde{\sigma}_p(\Delta N)$ for the large ($1.0\;\mu{\rm m}^2$) dot, averaged
over the six traces in Fig.\ 2 (c), at 85 mK (filled squares) and 170 mK (open
squares). Inset: same data for peak heights.  (b) Same as in (a) for the small ($0.3\;\mu{\rm
m}^2$) dot at 45 mK (closed circles) and 300 mK (open circles).} \label{fig3} \end{figure}

\begin{figure}[bth] \epsfxsize=\columnwidth \epsfbox{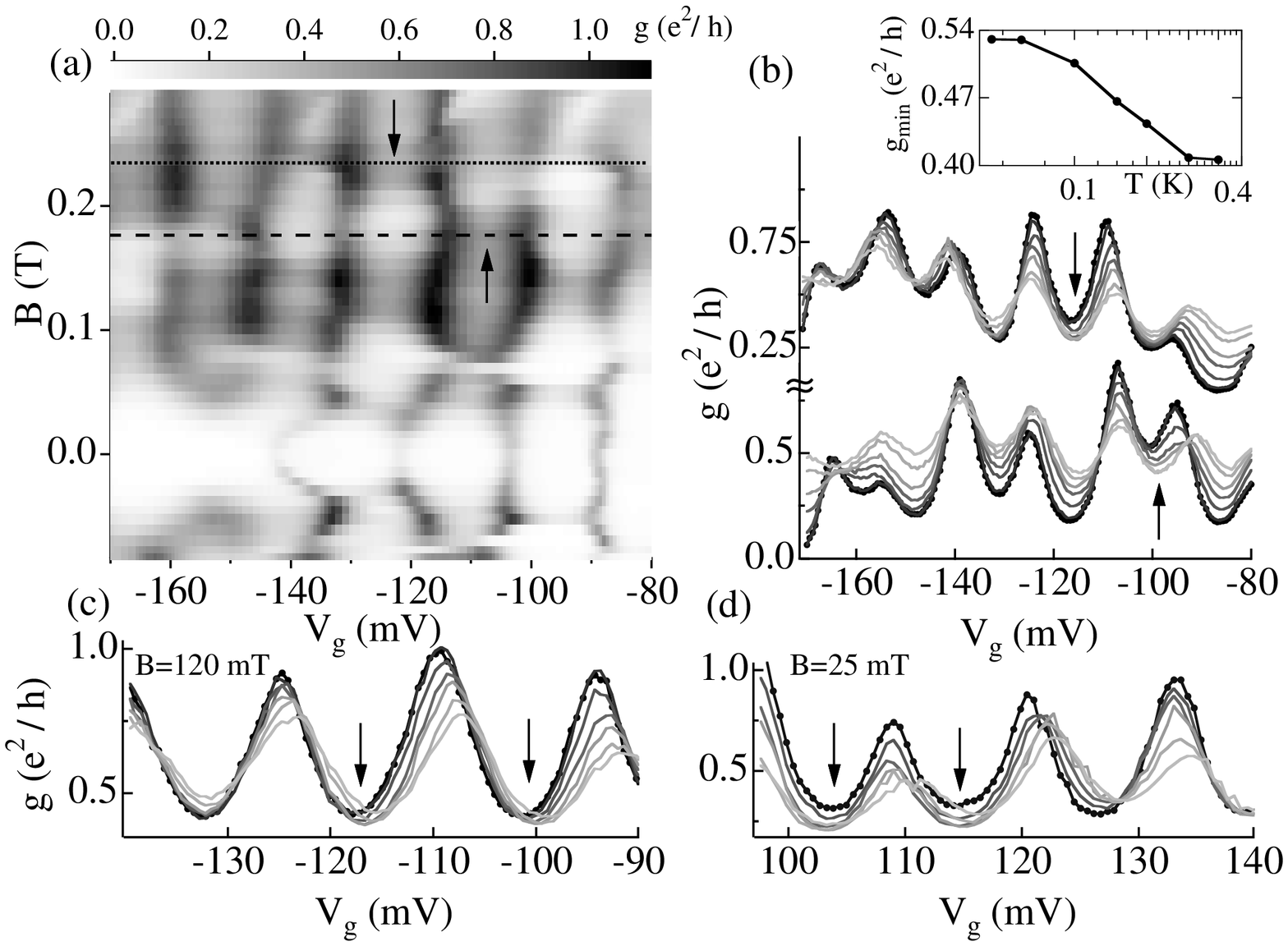}
\vspace{1 in}
\caption{(a) Conductance (grayscale) as a function of gate voltage $V_g$
and magnetic field $B$ for the small ($0.3\;\mu{\rm m}^2$) dot at 45 mK. (b)
Conductances of the two slices in (a) over temperatures from 45
mK (darker, with markers) to 400 mK (lighter). Adjacent valleys with anomalous temperature
dependence are marked with arrows. Inset: Conductance in the valley marked by
the up arrow. (c,d) Two instances of frequently observed anomalous temperature dependence in
adjacent valleys at fixed B.}
\label{fig4}
\end{figure}


\begin{thebibliography}{10}

\bibitem{CB} Single Charge Tunneling, edited by H. Grabert and M. H. Devoret
(Plenum, New York, 1992); L. P. Kouwenhoven {\it et~al.} in Mesoscopic
Electron Transport, edited by L. L. Sohn, L. P. Kouwenhoven, and G. Sch\"on
(Kluwer, Dordrecht, 1997).

\bibitem{JalabertChangFolk} R. A. Jalabert, A. D. Stone, and Y. Alhassid, Phys.
Rev. Lett. {\bf 68}, 3468 (1992); A. M. Chang {\it et~al.}, Phys. Rev. Lett.
{\bf 76}, 1695 (1996); J. A. Folk {\it et~al.}, Phys. Rev. Lett. {\bf 76}, 1699
(1996).

\bibitem{AleinerCot} I. L. Aleiner and L. I. Glazman, Phys. Rev. Lett. {\bf 77},
2057 (1996).

\bibitem{SaraValleys} S. M. Cronenwett {\it et~al.}, Phys. Rev. Lett. {\bf 79},
2312 (1997).

\bibitem{Aleiner} I. L. Aleiner and L. I. Glazman, Phys. Rev. B {\bf 57}, 9608
(1997).

\bibitem{SaraFrying} S. M. Cronenwett {\it et~al.}, Phys. Rev. Lett.
{\bf 81}, 5904 (1998).

\bibitem{KondoTheory} L. I. Glazman and M. E. Raikh, JETP Lett. {\bf 47}, 452
(1988); T. K. Ng and P. A. Lee, Phys. Rev. Lett. {\bf 61}, 1768 (1988); S.
Hershfield, J. H. Davies, J. W. Wilkins, Phys. Rev. Lett. {\bf 67}, 3720 (1991);
Y. Meir, N. S. Wingreen, P. A. Lee, Phys. Rev. Lett. {\bf 70}, 2601 (1993); N.
S. Wingreen and Y. Meir, Phys. Rev. B {\bf 49}, 11040 (1994); J. K\"onig {\it
et~al.}, Phys. Rev. B {\bf 54}, 16820 (1996).

\bibitem{Goldhaber} D. Goldhaber-Gordon {\it et~al.}, Nature {\bf 391}, 156
(1998).

\bibitem{SaraKondo} S. M. Cronenwett {\it et~al.}, Science {\bf 281}, 540
(1998).

\bibitem{Heinzel} T. Heinzel {\it et~al.}, Phys. Rev. B.
{\bf 52}, 16638 (1995).

\bibitem{Berman} D. Berman {\it et~al.}, Phys. Rev. Lett.
{\bf 82}, 161 (1998).


\bibitem{Kaminski} A. Kaminski, I. L. Aleiner, L. I. Glazman, Phys. Rev. Lett.
{\bf 81}, 685 (1998).

\bibitem{Brouwer} P. W. Brouwer and I. L. Aleiner, Phys. Rev. Lett. {\bf 82} 390 (1999).

\bibitem{FlensbergMatveev} K. Flensberg, Phys. Rev. B {\bf 48}, 11156 (1993); K.
A. Matveev, Phys. Rev. B {\bf 51}, 1743 (1995); A. Furusaki and K. A. Matveev,
Phys. Rev. Lett. {\bf 75}, 709 (1995).

\bibitem{Golubev} D. S. Golubev {\it et~al.}, Phys. Rev. B {\bf 56}, 15782
(1997).

\bibitem{Patel} R. O. Vallejos, C. H. Lewenkopf, and E. R. Mucciolo, Phys. Rev. Lett. {\bf 81},
677 (1998); S. R. Patel {\it et~al.}, Phys. Rev. Lett. {\bf 81}, 5900 (1998). 

\bibitem{SamPeakSpacings} S. R. Patel {\it et~al.}, Phys. Rev. Lett. {\bf 80},
4522 (1998).

\bibitem{SivanSimmel} U. Sivan {\it et~al.}, Phys. Rev. Lett. {\bf 77}, 1123
(1996); F. Simmel {\it et~al.}, Europhys. Lett. {\bf 38}, 123 (1997).

\bibitem{Berkovits} R. Berkovits, Phys. Rev. Lett. {\bf 81}, 2128 (1998).


\bibitem{drift} To further reduce experimental noise, all data were taken at
positive and negative $B$ values, and a best-fit straight line was subtracted
from the peak position to remove small amounts of drift.

\bibitem{RMTAAK} M. L. Mehta, {\it Random Matrices} (Academic Press, New York,
1967); B. L. Altshuler and B. I. Shklovskii, Sov. Phys. JETP {\bf 64} 127
(1986).

\bibitem{PatelMaurer} S. R. Patel, S. M. Maurer, C. M. Marcus (unpublished).

\bibitem{renorm} L. I. Glazman and K. A. Matveev, Sov. Phys. JETP {\bf 71}, 1031
(1990); H. Schoeller and G. Sch\"on, Phys. Rev. B {\bf 50}, 18436 (1994); J.
K\"onig, H. Schoeller, and G. Sch\"on, Europhys. Lett. {\bf 31}, 31 (1995); L.
W. Molenkamp, K. Flensberg, and M. Kemerink, Phys. Rev. Lett. {\bf 75}, 4282
(1995); P. Joyez {\it et~al.} Phys. Rev. Lett. {\bf 79}, 1349 (1997).

\bibitem{Fluctuations} Y. M. Blanter {\it et~al.}, Phys. Rev. Lett. {\bf 78},
2449 (1997); R. Berkovits {\it et~al.}, Phys. Rev. B {\bf 55}, 5297 (1997).


\bibitem{Duncan} D. R. Stewart {\it et~al.}, Science {\bf 278}, 1784 (1997).

\bibitem{Inoshita} T. Inoshita, Y. Kuramoto and H. Sakaki, Superlattices and
Microstructures {\bf 22}, 75 (1997).

\end{thebibliography}
\end{document}